\documentclass[twocolumn,10pt,conference]{IEEEtran}
\usepackage{graphicx}
\usepackage{amsmath} 
\usepackage{subfigure}
\usepackage{hyperref}
\usepackage{breakurl}

\newtheorem{thrm}{Theorem}[section]
\newtheorem{myprop}[thrm]{Proposition}

\title{Two Algorithms for Network Size Estimation for Master/Slave Ad Hoc Networks}
\author{Redouane Ali, Suksant Sae Lor, Miguel Rio\\
University College London\\
\{r.ali, s.lor, m.rio\}@ee.ucl.ac.uk}
\begin{document}

\maketitle

\begin{abstract}
This paper proposes an adaptation of two network size estimation methods: random tour and gossip-based aggregation to suit  
master/slave mobile ad hoc networks. We show that it is feasible to accurately estimate the size of ad hoc networks when 
topology changes due to mobility using both methods. The algorithms were modified to account for the specific constraints 
of master/slave ad hoc networks and the results show that the proposed modifications perform better on these networks than 
the original protocols. Each of the two algorithms presents strengths and weaknesses and these are outlined in this paper. 
\end{abstract}

\section{Introduction}
Mobile ad hoc networks are self-configuring networks of mobile devices that do not rely on any underlying infrastructure and are characterised by frequent changes in topology. 
It is common for these networks to be arranged in clusters for increased quality of service and ease of management \cite{clustering_survey}. 
A particular energy-efficient clustering mechanism is to arrange nodes in pico-networks, in which the clusterhead assumes the role of master while the remaining nodes within the cluster act as slaves \cite{master_slave}. A master performs general cluster management such as transmission scheduling \cite{bluetooth} or acts as WAN/LAN gateway \cite{802_master_slave}. In a master/slave setting, slave nodes can only communicate with each other via the master node.

Network size estimation is an instance of applications that must account for the dynamic nature of ad hoc networks. Since it 
is not practically feasible to maintain an accurate topology view of the network due to constant change, 
some applications rely on the knowledge of its approximate size instead. 


There have been many proposals for estimating the size of large peer-to-peer networks; Two major techniques can be 
distinguished: Gossip-based aggregation\cite{ref21,ref22} and random walk-based methods \cite{ref20,ref23}. In the former approach, nodes average a value each time they 
communicate with each other so that if the process is repeated enough times, nodes would have some average, from which 
the size of the network can be infered. It was shown in \cite{ref24} that for sizes of up to 1,000,000 nodes, 
gossip-based aggregation obtains an accurate estimate of the network size after just 40 rounds. Here a round is a repetition 
of the aggregation after a certain time has elapsed (called an \textit{epoch}). Random walk methods, on the other hand, 
trade accuracy for lower cover time. These methods rely on passing a token around, gathering network measurements and estimating the size 
of the network upon return of the token to the initiator. The 
work presented herein adapts these two methods to suit the particular constraints of master/slave ad hoc networks, taking into account the effect of node mobility.

\section{Random Tour Method}
The random walk estimation method employed herein is based on the work described in \cite{ref20}, which 
proposes a random tour technique for estimating the size of large peer-to-peer networks. In this method an originator node, 
$i$, gathers information along the walk and estimates the size of the network upon return of the walk to this originator. 

The originator, $i$, initialises a counter $X$ to $\frac{1}{d_i}$, where $d_i$ is the degree of node $i$, 
and forwards the message to one of its randomly selected neighbours. Upon receipt of the message, a node $j$, ($j\neq i$), 
increments the counter by $\frac{1}{d_j}$, $d_j$ being the degree of node $j$, and forwards it to one of its immediate 
neighbours, again chosen randomly.
When the message returns back to the originator, this node performs the estimate of the size of the network as: 
$\Phi=d_iX$

Random walk methods also allow gathering other network statistics \cite{ref25} such as node degrees and roles, which are 
key ingredients in the proposed algorithm. The Random Tour method \cite{ref20} can, therefore, be adapted to suit the 
specific attributes of master/slave settings in order to yield more accurate estimates of the size of the network. 

Such a topology forms a bipartite graph with vertices of colours \textit{m}\ and \textit{s}\ for masters and slaves 
respectively. Let $G(V,E)$ define a master/slave ad hoc network in which the set of vertices $V(G) = M(G) \cup S(G)$, where 
$M(G)$ and $S(G)$ are the set of master nodes and slave nodes respectively. We distinguish two subgroups in 
$S(G)=PS(G) \cup P(G)$ where $PS(G)$ is the set of all s-coloured vertices whose degree is exactly one and $P(G)$ is the 
set of all s-coloured vertices whose degree is greater than one. We shall refer to nodes in $M(G)$ as \textit{\bf{master}} 
nodes, nodes in $P(G)$ as \textit{\bf{PMP}} nodes and nodes in $PS(G)$ as \textit{\bf{pure slave}} nodes.\\
Let $N=|V(G)|$, $M=|M(G)|$, $S=|S(G)|$, $P=|P(G)|$ and $PS=|PS(G)|$. Hence we have $N=M+S$ and $S=P+PS$.
We define 
$\overline{d_m}=\frac{\sum_{v\in M(G)} d(v)}{M}$ 
and 
$\overline{d_s}=\frac{\sum_{k\in S(G)} d(k)}{S}$ 
as the average node degrees of \textit{m}-coloured and \textit{s}-coloured nodes in $V(G)$ respectively. Similarly we 
define the average degree of vertices in subgroup $P(G)$ as
$\overline{d_p}=\frac{\sum_{n\in P(G)} d(n)}{P}$ 

\begin{myprop}
The total number of nodes ($N$) in a connected component within the network is given by:
\begin{center}
$N = M\cdot (\overline{d_m}+1) - P\cdot (\overline{d_p} - 1)$
\label{proposition}
\end{center}
\end{myprop}

\begin{IEEEproof}
 Because the graph is bi-partite, every egde in $E(G)$ is incident in $M(G)$ and in $S(G)$; therfore:
\begin{center}
$e(G)=\sum_{v\in M(G)} d(v) = \sum_{k\in S(G)} d(k)$
\end{center}
where $e(G)=\|E(G)\|$ is the number of edges in $G$.\\
From the previous definitions we have:
\begin{equation}
 S=M\cdot d_m-S\cdot (\overline{d_s}-1)
 \label{eq1}
\end{equation}
\begin{equation}
 \sum_{k\in S(G)}d(k)=\sum_{n\in P(G)}d(n)+\sum_{i\in PS(G)} d(i)
 \label{eq2}
\end{equation}
\begin{equation}
 \overline{d_s}=1\Rightarrow\sum_{i\in PS(G)}d(i)=PS
 \label{eq3}
\end{equation}
it follows from (\ref{eq1}), (\ref{eq2}) and (\ref{eq3})
\begin{center}
 $S\cdot (\overline{d_s}-1)=P\cdot (\overline{d_p}-1)\Rightarrow S=M\cdot \overline{d_m}-P\cdot (\overline{d_p}-1)$
\end{center}
Consequently we get 
\begin{center}
 $N = M\cdot (\overline{d_m}+1) - P\cdot (\overline{d_p} - 1)$
\end{center}
\end{IEEEproof}

The function of this estimator is to obtain an accurate size of the network, the exact value of which is $N$. The network size estimation process is performed by each master node, which upon completion broadcasts the 
estimate to its neighbours so that every node in the network would have the estimate. 

At any non-initial state $\sigma$, a node $j$, 
($j\neq i$) selects one of its outgoing links at random with probability $p_j=\frac{1}{d_j}$ and forwards the estimate 
message to the corresponding node. The estimate message is the tuple \{($N_m$,$D_m$),($N_p$,$D_p$)\}, where $N_m$ 
is a counter for the number of master nodes discovered so far and $N_p$ is a counter for the number of PMP nodes. $D_m$ and 
$D_p$ represent the respective cumulative counts of the degrees of masters and PMP nodes encountered so far during 
the walk. The estimate message is initialised at the originating master $i$ to \{(1,$d_i$),(0,0)\} to indicate that, thus 
far, there has been only one master node discovered with degree $d_i$ and no PMP nodes. Upon receipt of the message, a receiving node increments the counter $N_m$ by $1$ and adds its degree to the counter $D_m$ if it is a master 
node or updates $N_p$ and $D_p$ in case of a PMP; the message remains unchanged if the node is a pure slave.
When   the   initiating   node   receives   the message back, i.e. when the tour is completed, that same node computes the average masters and PMP nodes' degrees as $d_m=\frac{D_m}{N_m}$ and $d_p=\frac{N_p}{D_p}$ respectively then 
estimates the size of the network according to proposition \ref{proposition}. The operation of the network size estimation is similar for 
both static and dynamic cases in the absence of node crashes; this is when node movement changes the topology but the 
network remains connected. In \cite{ref26} the authors give the argument that a change in topology might be considered as a 
probability to chose the next node to forward the message to and prove that the walk would complete in $O(n\log n)$ for a 
complete graph. However, in our case the graph is not complete and the topology is governed by strict connectivity rules, 
therefore we cannot assume that the algorithm will eventually converge in such scenarios. Furthermore, neither \cite{ref20} 
nor \cite{ref26} consider the case when it is impossible to forward the message (token) if, for instance, the 
node that currently holds the token crashes before forwarding it or moves out of range of the next node to receive the 
message. To account for these facts, we propose the following modification to the random tour:

When a node, $i$, receives the token, it is marked as having been visited and forwards the message to one of its 
neighbours, say node $j$, selected randomly with probability $\frac{1}{d_i}$. Node $j$ then forwards the message to node 
$q$, again selected randomly, and sends an acknowledgement back to node $i$ indicating that is has forwarded the message. If node $i$ 
fails to hear the acknowledgement from node $j$, within a predefined timeout, it selects another node to forward the 
message to. When node $i$ does not find any unmarked neighbouring node, it assumes that the message cannot travel 
any further and returns it to the initiator using an ad hoc routing protocol such as Ad hoc On-demand Distance Vector (AODV) or Dynamic Source Routing (DSR). This ensures that tokens are not lost or remain in an infinite loop without ever returning to the originator.  

\begin{figure}[h!]
\begin{center}
\includegraphics[height=2.5in,width=3.1in]{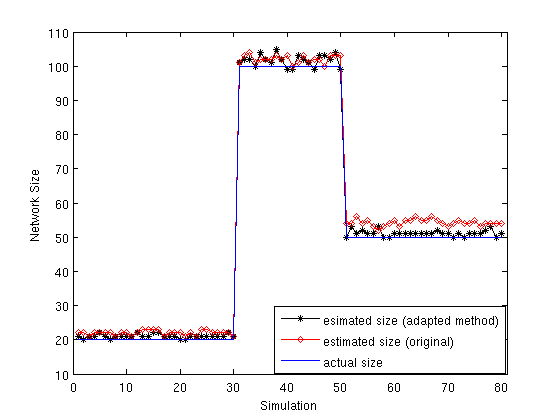}
\caption{Network Size Estimations (Random Tour Method)}
\label{random_tour}
\end{center}
\end{figure}

Figure \ref{random_tour} shows simulation results of network size estimations generated with the random tour technique described in this 
paper, compared with the method proposed in \cite{ref20} and the actual sizes of various randomly generated networks 
with different assumptions of node mobility. 

It can be seen from Figure \ref{random_tour} that the proposed adaptation yields better estimates, which 
proves the adequacy of this method. However, an apparent difference between the method employed herein and the original 
approach, given in \cite{ref20}, is that the former presents less bias than the latter. Another observation is the fact that 
the estimated size is not always exact but that is expected from such an approach.

\section{Gossip-based Aggregation}
In gossip-based aggregation algorithms, an initiating node sets a variable, $avg$, to $1$ with all other nodes having this 
variable set to $0$. The initiator, $i$, chooses one of its outgoing links randomly with probability $\frac{1}{d_i}$ and 
exchanges information with the corresponding node, $p_{next}$. Both nodes then average their $avg$ variable as 
$\frac{avg_i+avg_{p_{next}}}{2}$ and store the value. If this process is repeated infinitely then every node in the network 
will have its $avg$ variable set to exactly $\frac{1}{N}$, where $N$ is the number of nodes in the network.
In a master/slave setting, however, we can take advantage of the topological characteristics of such networks. Nodes are 
arranged in pico-networks (or clusters) and coordinated by a master node (or cluster head). In the proposed protocol,
the master collects the current averages from all slaves, performs the averaging then broadcasts the results to every node 
in the pico-network. If this process is repeated for a sufficiently large number of rounds then every node in the connected 
component will be able to accurately derive the size of the network ($N$). Here a \textit{round} is defined as the process of 
collecting, averaging and re-distributing the information that each master node performs.  

An obvious advantage of this technique is that node mobility does not deteriorate the precision of the estimate, as 
regardless of nodes' location in the network, the sum of all the averages will always converge to unity, that is: 
$\sum_{i\in{V(G)}}{avg_i}=1$, from which we can infer by logical induction
\begin{equation}
 \forall{i}\in{V(G)}:\lim_{k\rightarrow \infty}\left(\frac{1}{avg_i(k)}\right)=||V(G)||
\end{equation}
where $V(G)$ is the set of nodes in the connected component, $avg_i$ is the average at node $i$ and $k$ is the number of 
aggregation rounds.

\begin{figure}[h!]
\begin{center}
\includegraphics[height=2.5in,width=3.2in]{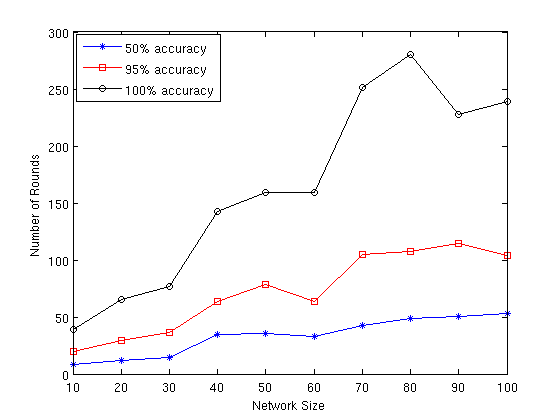}
\caption{Gossip-based aggregation}
\label{aggr}
\end{center}
\end{figure}

Because the number of rounds needed to find the exact network size can, in theory, be infinite, in practice algorithms are 
considered converged if the estimate reaches a certain precision level \cite{ref21}. Since the protocol is applied to 
ad hoc networks, with much fewer nodes than large peer to peer systems, the algorithm does converge towards the exact 
network size in a finite time. Figure \ref{aggr} shows simulation results for the number of rounds taken to find the size estimate 
of various randomly generated master/slave networks, for different estimation precisions and under various assumptions of 
node mobility conditions.

It is to note, from Figure \ref{aggr}, that the quality of the estimate does not follow a linear trend with the number of rounds. 
This indicates that for very accurate estimates, the algorithm needs to run for a large number of rounds. However, the 
technique proposed in this paper requires far less time than does the gossip-based aggregation method, proposed in 
\cite{ref22}. Table \ref{tbl} shows a comparison between the original and the adapted algorithm for 100\% precision level for 
different network sizes. 

\begin{table}[h!]
\caption{Comparison between original and adapted gossip-based aggregation algorithm}
\centering
\begin{tabular}{|c|c|c|c|}
\hline
Network size (number of nodes). & 20 & 60 & 100 \\
\hline
Number of rounds (original algorithm) & 2544 & 5035 & 7521\\
\hline
Number of rounds (proposed algorithm) & 132 & 232 & 277.\\
\hline
\end{tabular}
\label{tbl}
\end{table}
The reason for this significant difference stems from the fact that in each round, the adapted method produces the average of all participating nodes in the cluster, whereas in the original approach the aggregation concerns a single master-slave pair.

\section{Conclusion}
This paper has adapted two network size estimation methods to suit the master/slave architecture and has shown that both can 
accurately estimate the size of the network. The gossip-based aggregation method exhibits better accuracy than the random 
tour method but at the expense of longer convergence times. On the other hand, the random tour algorithm allows the 
collection of other network statistics such as average node degree that could be useful to other applications. Both methods cope with random topology 
changes, which makes them suitable for mobile ad hoc networks.

\bibliographystyle{IEEEtran}
\bibliography{references}
\end{document}